\documentclass[prb,aps]{revtex4}
\usepackage{graphicx}

\begin{document}

\title{Giant spin accumulation in silicon nonlocal spin-transport devices}
\vspace*{3mm}
\author{A. Spiesser$^1$, H. Saito$^1$, Y. Fujita$^2$, S. Yamada$^2$, K. Hamaya$^{2,3}$, S. Yuasa$^1$ and R. Jansen$^1$}
\affiliation{$^1\,$Spintronics Research Center, National Institute of Advanced Industrial Science and Technology (AIST), Tsukuba, Ibaraki, 305-8568, Japan.\\
$^2\,$Department of Systems Innovation, Graduate School of Engineering Science, Osaka University, Toyonaka 560-8531, Japan.\\
$^3\,$Center for Spintronics Research Network, Graduate School of Engineering Science, Osaka University, Toyonaka 560-8531, Japan.}

\begin{abstract}
Although the electrical injection, transport and detection of spins in silicon have been achieved, the induced spin accumulation was much smaller than expected and desired, limiting the potential impact of Si-based spintronic devices. Here, using non-local spin-transport devices with an n-type Si channel and Fe/MgO magnetic tunnel contacts, we demonstrate that it is possible to create a giant spin accumulation in Si, with the spin splitting reaching 13 meV at 10 K and 3.5 meV at room temperature. The non-local spin signals are in good agreement with a numerical evaluation of spin injection and diffusion that explicitly takes the size of the injector contact into account. The giant spin accumulation originates from the large tunnel spin polarization of the Fe/MgO contacts (53 \% at 10 K and 18 \% at 300 K), and the spin density enhancement achieved by using a spin injector with a size comparable to the spin-diffusion length of the Si. The ability to induce a giant spin accumulation enables the development of Si spintronic devices with a large magnetic response.\end{abstract}

\maketitle

\section{INTRODUCTION}
\indent Exploiting the spin degree of freedom in semiconductors enables the development of novel devices and systems with characteristics and functionalities that are distinct from those of traditional charge-based semiconductor electronics\cite{zutic,fabianacta,awschalom,fertnobel,sugaharanitta,jansennmatreview}. These semiconductor spintronic devices rely on the ability to induce a non-equilibrium spin density (i.e., a spin accumulation) in the otherwise non-ferromagnetic semiconductor, and to detect and manipulate it, all in an efficient manner. The creation of a spin accumulation in semiconductors is generally achieved by driving an electrical current from a ferromagnetic (FM) tunnel contact into the semiconductor. This produces a spin current into the semiconductor due to spin-polarized tunneling. Ferromagnetic tunnel contacts are also used to convert a spin accumulation into a detectable (charge) voltage. Not surprisingly, much of the research has focused on mainstream semiconductors, such as silicon, which are compatible with existing electronics and also exhibit a sufficiently large spin lifetime $\tau_s$ (of the order of nanoseconds\cite{jansennmatreview}).\\
\indent In order to establish the presence of a spin accumulation in a semiconductor and obtain quantitative information, one generally uses a so-called non-local measurement geometry\cite{johnson,jedema,lou}. In this geometry, one ferromagnetic contact is used as injector to induce a spin accumulation in the semiconductor channel, and the spin accumulation is detected using a second ferromagnetic electrode placed close to the point of injection at a distance comparable to or smaller than the characteristic spin-transport length (the spin-diffusion length, given by $L_{SD}$=$\sqrt{D\,\tau_s}$, where $D$ is the diffusion constant). Indeed, using such non-local devices, the electrical injection, transport and detection of spins in heavily-doped n-type silicon have been achieved\cite{erve,erveieee,suzukiAPEX2011,shiraishi,sasaki,toshibanonlocal,ishikawaPRB2017,toshiba2017}, including at room temperature. Unfortunately, the induced spin accumulation was very small and the detected spin signals (in the $\mu$V range) are about two orders of magnitude smaller than expected. Although the reason has not been clearly identified, within the standard theory for spin injection and diffusion\cite{jedema,lou,fert,maekawa}  the small spin accumulation translates into a small tunnel spin polarization of only 5 - 10 \% for the Fe/MgO tunnel contacts used for spin injection and detection\cite{erve,erveieee,suzukiAPEX2011,shiraishi,sasaki,toshibanonlocal,ishikawaPRB2017,toshiba2017}. However, much larger values (50\% or higher) are expected for crystalline Fe/MgO(001) tunnel contacts that are notorious for their large tunnel spin polarization \cite{parkin,yuasa,jiang} arising from symmetry-based spin filtering\cite{butler,mathon}.\\
\indent The issue of the small tunnel spin polarizations obtained so far has a technological and a scientific aspect. In order to design low-power spintronic devices and circuits, one naturally needs a large magnetic response, for which the efficient generation of a substantial spin accumulation is indispensable. The inability to create a large spin accumulation has thus far seriously limited any potential impact that Si-based spintronic devices might have. From a scientific point of view, the question is why the spin accumulation and the tunnel spin polarization are only small. Is this simply because the quality of the Fe/MgO contacts on Si has hitherto been insufficient? Or is there a more fundamental reason, namely, that coherent tunneling and the resulting symmetry-based spin filtering, which are the origin of the large tunnel magnetoresistance in metal (Fe/MgO/Fe) tunnel junctions, are not applicable to Fe/MgO contacts when fabricated on Si? Since the electronic structure of Si is not the same as that of Fe-based metallic ferromagnets, it is not at all obvious that the same symmetry-based spin filtering is applicable to Fe/MgO/Si tunnel junctions.\\
\indent Here, we demonstrate that it is possible to electrically create a giant spin accumulation in Si. The spin accumulation, characterized by a spin splitting $\Delta\mu$ of the electrochemical potential, reaches values as large as 13 meV at 10$\,$K and 3.5 meV at room temperature. The spin-valve and Hanle data obtained in non-local spin-transport devices with an n-type Si channel and Fe/MgO magnetic tunnel contacts are in good agreement with the theory for spin injection and spin diffusion\cite{fert,maekawa}, from which we extract a large tunnel spin polarization of the Fe/MgO contacts (53 \% at 10 K and 18 \% at 300 K). Also, we experimentally confirm an inherent, yet hitherto untested, aspect of the theory for spin injection and diffusion in non-magnetic materials, namely, that the spin density in the non-magnetic channel is enhanced when the lateral size of the spin injector contact is increased relative to the spin-diffusion length of the channel material. The observation of large spin signals, which amount to an improvement by about two orders of magnitude, demonstrate that it is possible to obtain a large tunnel spin polarization in Fe/MgO contacts on Si, and also eliminates an obstacle (small magnetic response) that prevents the technological impact of silicon spintronic devices.\\

\section{Results}
\subsection{Growth of Fe/MgO tunnel contacts on Si} The device fabrication starts with the growth of Fe/MgO tunnel contacts by molecular beam epitaxy (MBE) onto the Si substrate containing a 70 nm-thick epitaxial n-type Si(001) channel doped with phosphorous at a density of 2.7 $\times$ 10$^{19}$ cm$^{-3}$ (see Appendix A for details). Prior to MgO deposition, the {\it in-situ} reflection high-energy electron diffraction (RHEED) patterns of the Si surface showed intense and sharp streaks (Fig. 1a), which indicates a clean and smooth Si surface. The RHEED patterns correspond to a c(2 $\times$ 4) reconstruction, which has been ascribed to buckled dimers\cite{chadi}. After the growth of a 2 nm-thick MgO layer and a 10 nm-thick Fe layer, the RHEED images exhibit spotty patterns corresponding to crystalline MgO(001) (Fig. 1b) and bcc-Fe(001) (Fig. 1c), respectively. Transmission electron microscopy (TEM) reveals a good morphology of the Fe/MgO/Si structure, with a flat and continuous MgO tunnel barrier and sharp interfaces (Fig. 1d). The epitaxial MgO has a reasonable degree of crystallinity whereas the Fe layer is single crystalline, as can be seen in the high-resolution cross-sectional TEM image (Fig. 1e). Defects are also present in the MgO layer, as expected from the lattice mismatch between MgO and Si (3.9 \% for a cube-on-cube growth with a unit cell ratio of 4:3).

\begin{figure}[htb]
\centering
\includegraphics*[width=110mm]{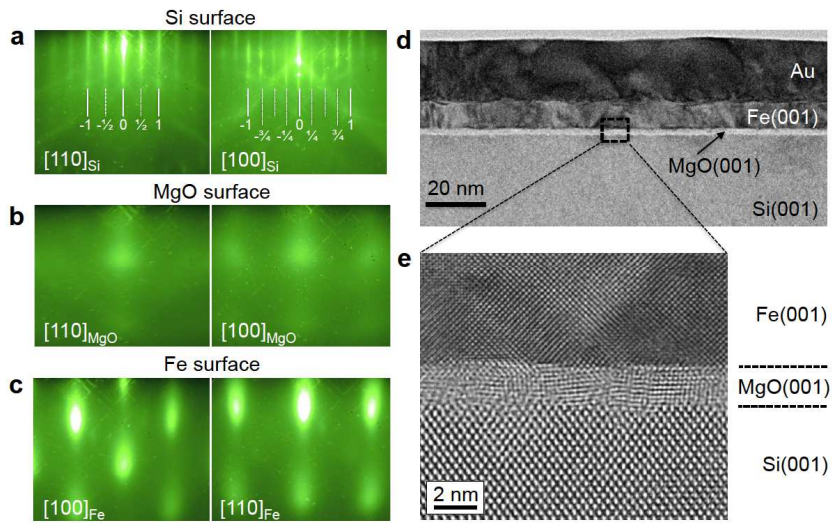}
\caption{Structural characterization of the Si/MgO/Fe tunnel contact. RHEED patterns of (a) the Si surface having a c(2 $\times$ 4) reconstruction after {\em in situ} annealing at 700$^{\circ}$C, (b) the MgO(001) layer deposited at 300$^{\circ}$C and (c) the Fe(001) layer deposited at 200$^{\circ}$C. In each case the patterns along the [100] and [110] azimuths are shown. Note that the azimuth labels for the Fe are interchanged, because the Fe lattice is rotated by 45$^{\circ}$ with respect to the MgO and Si lattices. (d) Low magnification and (e) high-resolution cross-sectional TEM images of the junction.}
\label{fig1}
\end{figure}

\subsection{Spin transport in Si non-local devices} The creation of spin accumulation in the Si is probed in non-local devices that consist of a Si channel, shaped in the form of a strip, with 4 electrical contacts (Fig. 2a). The two outer non-magnetic Au/Ti contacts serve as reference contacts, whereas the two central contacts are Fe/MgO magnetic tunnel contacts with a separation $d$. A charge current across the interface of one of these FM contacts is accompanied by a spin current into the Si and thereby induces a spin accumulation in the Si channel. It decays exponentially in both directions by spin diffusion on a length scale set by $L_{SD}$, and is detected by probing the voltage across the second ferromagnetic tunnel contact. The detected non-local voltage $V_{NL}$ is given by $P_{det}\,\Delta\mu /2e$, where $P_{det}$ is the tunnel spin polarization of the detector tunnel contact, $\Delta\mu$ the spin accumulation under it and $e$ is the electron's charge. The $V_{NL}$ changes sign when either the injector magnetization is reversed ($\Delta\mu$ changes sign) or the detector magnetization is reversed ($P_{det}$ changes sign). Therefore, when an external magnetic field is applied and the relative magnetization of injector and detector is changed between the parallel (P) and anti-parallel (AP) states, the non-local voltage changes sign. This is indeed what is observed in the measurement when the applied magnetic field is in-plane (B$_Y$) along the long axis of the FM contacts (Fig 2b, left panel). The $V_{NL}$ has a value of about $+$0.5 mV when the magnetizations of the two FM contacts are parallel, and a value of about $-$0.5 mV for the AP state, and sharp transitions when the magnetization of either the injector or detector contact is reversed. We confirm that this typical spin-valve signal is indeed due to spin accumulation in the Si channel by performing non-local Hanle measurements (Fig. 2b, right panel) with the magnetic field (B$_Z$) applied perpendicular to the magnetization and thus to the injected spins. This causes spin precession and a reduction of $V_{NL}$ from its maximum value at B$_Z$=0, to zero at large enough magnetic field for which the spin accumulation is completely suppressed. As expected, the Hanle signal has the opposite sign for the P and AP state of the magnetizations of injector and detector, and the signal magnitude (0.5 mV) is consistent with the spin-valve data. This proves unambiguously that the signal is genuine and due to spin accumulation in the Si channel and the corresponding transport of spins from the injector to the detector.

\begin{figure}[htb]
\centering
\includegraphics*[width=97mm]{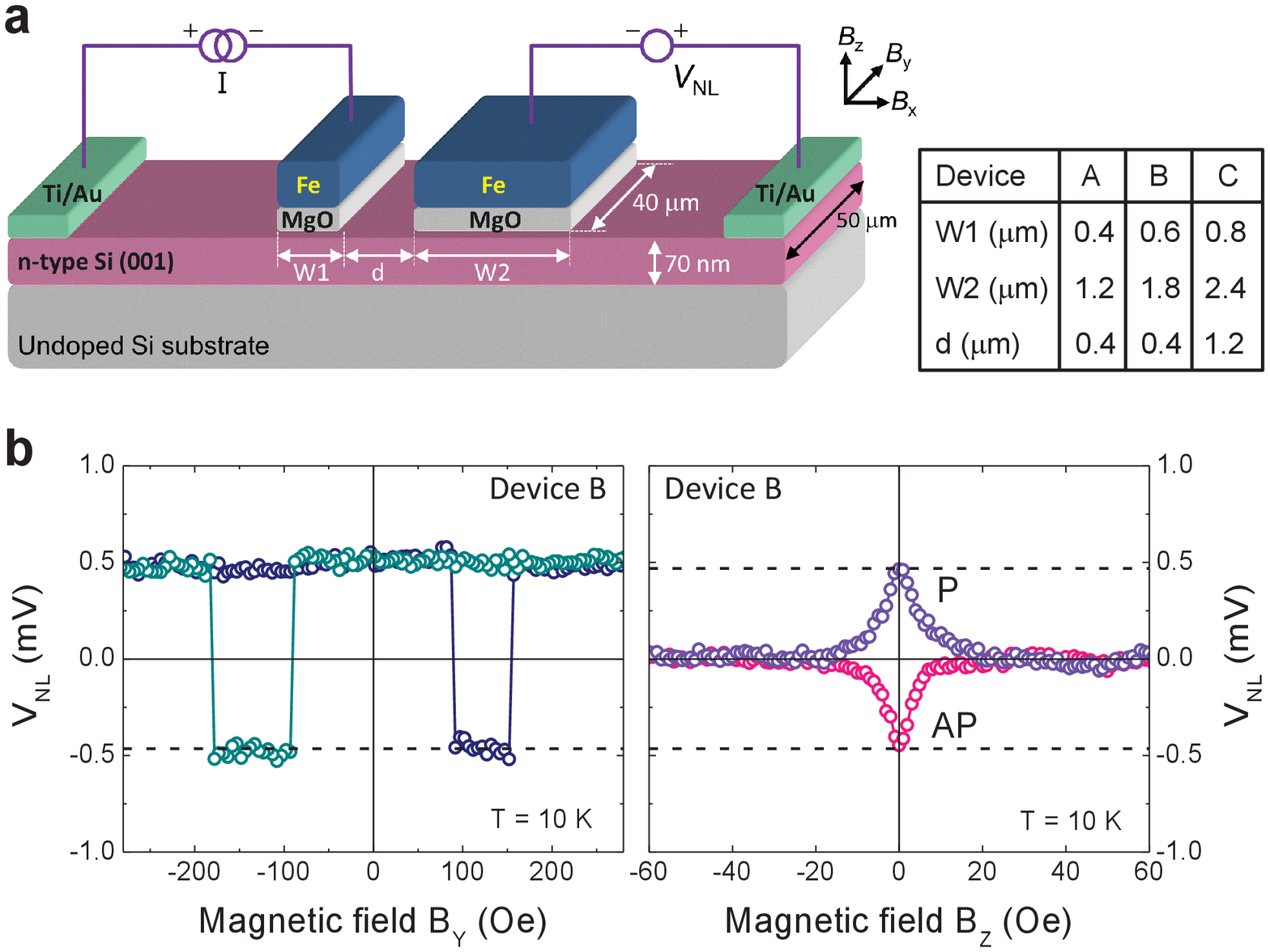}\hspace*{5mm}\includegraphics*[width=70mm]{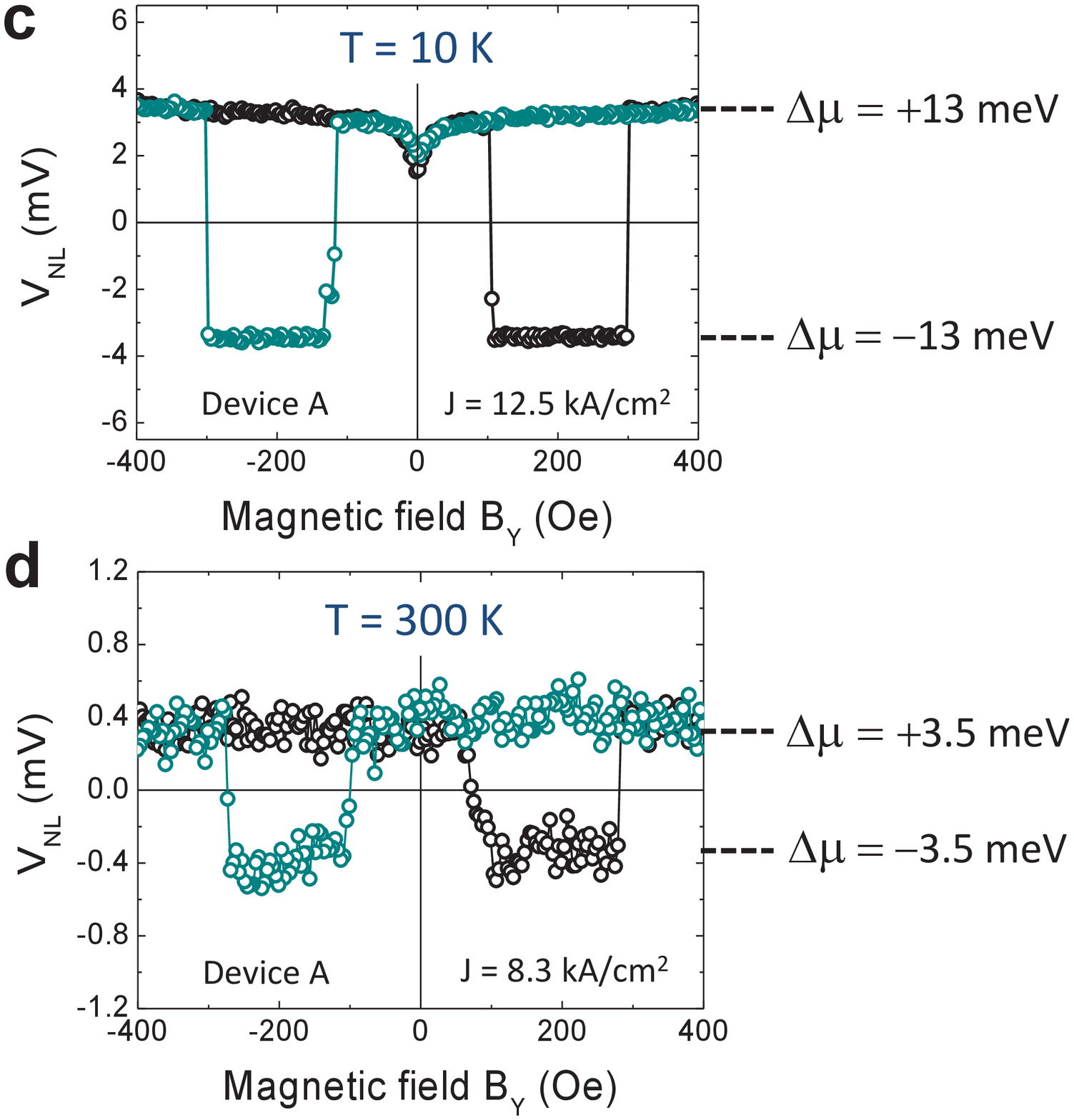}
\caption{Spin transport in Si non-local device and giant spin accumulation. (a) Schematic layout of the non-local device with dimensions indicated. (b) Non-local spin signals $V_{NL}$ measured on device B in spin-valve geometry and Hanle geometry with the external magnetic field applied, respectively, in-plane (B$_Y$) along the long axis of the FM contacts, or perpendicular to the sample plane (B$_Z$). The narrow FM strip (0.6 $\mu$m) was used as injector, and the spin accumulation in the Si was detected using the wide FM contact (1.8 $\mu$m). The B$_Y$ is either swept from plus to minus (green symbols), or in the opposite direction (dark blue symbols). The wide (narrow) FM contact reverses its magnetization at a smaller (larger) value of B$_Y$. The injected current $I$ was +1 mA (current density $J$=+4.2 kA/cm$^2$, electrons flowing from FM into the Si) and an offset of about 3 mV was subtracted from the measured signals. T = 10 K. (c,d) Non-local spin-valve signals measured on device A at 10 K and at 300 K, using the wide FM strip (1.2 $\mu$m) as injector and the narrow FM strip (0.4 $\mu$m) as non-local detector. Indicated are the values of the spin accumulation under the detector contact (either positive or negative), extracted from $V_{NL} = P_{det}\,\Delta\mu\,/2e$ using $P_{det}$ is 53\% at 10 K and 18\% at 300 K. The $J$ was +12.5 kA/cm$^2$ ($I$=+6 mA) at 10 K and +8.3 kA/cm$^2$ ($I$=+4 mA) at 300 K. The origin of the cusp around zero field in (c) is not understood.}
\label{fig2}
\end{figure}

\subsection{Giant spin accumulation in silicon} Next, it is demonstrated that the spin accumulation can be very large. Fig. 2c shows the non-local spin-valve measurement obtained at a temperature (T) of 10 K on device A using the wider of the two FM contacts as the injector of spins and the narrow FM strip as the non-local detector. The current density across the injector interface was +12.5 kA/cm$^2$, which, for comparison, is 2 to 3 orders of magnitude smaller than what is typically used for spin-torque magnetization reversal in the metal tunnel junctions of magnetic random access memory\cite{mram1,mram2,mram3}. A characteristic spin-valve behavior is observed, but most strikingly, the non-local spin signal $V_{NL}$ reaches a magnitude of $+$/$-$ 3.5 mV. Such huge spin signals are unprecedented. The signal is converted into a spin accumulation via $V_{NL}$ = $P_{det}\,\Delta\mu /2e$ using the value of $P_{det}$ = 53\% that is determined below. We obtain $\Delta\mu$ = 13 meV. Thus, the spin accumulation in the Si channel is giant. But note that it is not unreasonably large, i.e., it is in line with what is expected for spin injection from Fe/MgO tunnel contacts with a reasonable tunnel spin polarization, given the device parameters and the current density used, as will be shown below. Equally important, when the temperature is increased to 300 K, a large spin signal still remains (Fig. 2d), with a corresponding spin accumulation of about 3.5 meV (using $P_{det}$ = 18\% at 300 K, see below). Thus, our results demonstrate that a giant spin accumulation can indeed be created in degenerately-doped Si, not only at low temperature, but at room temperature as well. In the next sections we will provide a precise description of the spin signals based on numerical calculations of the spin accumulation profile, which allows us to establish how the spin signal depends on various parameters and identify the main origin of the giant spin accumulation.\\

\subsection{Calculation of the spin accumulation profile} The spatial profile of the injected spin accumulation is obtained from the expression
for one-dimensional spin diffusion, spin precession and spin relaxation in a semiconductor\cite{zutic,fabianacta}. Integration over time $t$ and the size of
the injector contact $W_{inj}$ in the $x$-direction yields the spin accumulation at location $x$ in the Si channel produced by spins injected
from the injector contact between $x=-W_{inj}$ and $x=0$:
\begin{equation}
\Delta\mu (x) = 2e\,\,J\,P_{inj}\,r_{ch} \int_{-W_{inj}}^{0}\int_{0}^{\infty} \frac{1}{\tau_s}\,\frac{1}{\sqrt{4\pi Dt}}\exp\left(-\frac{(x-x_1)^2}{4Dt}\right)\,\cos\left(\frac{g\mu_B B_Z}{\hbar}\,t\right)\,\exp\left(-\frac{t}{\tau_s}\right) \,dtdx_1 \label{eqn1}
\end{equation}
where $P_{inj}$ is the tunnel spin polarization of the spin injector contact, $\mu_B$ the Bohr magneton, $g$ the electron $g$-factor and $B_Z$ the magnetic field perpendicular to the spins in case of a Hanle measurement \cite{note1}. The one-dimensional approach is justified since in our experiments the thickness of the Si channel ($t_{Si}$ = 70 nm) is much smaller than the spin-diffusion length $L_{SD}$ ($\sim$ 1-2 $\mu$m), so that the spin accumulation is essentially homogeneous in the z-direction perpendicular to the tunnel interface. The effective spin resistance of the channel with resistivity $\rho$ is then given by\cite{fert,jansensstreview} $r_{ch}=\rho\,L_{SD}(L_{SD}/t_{Si})$, which includes the geometric correction factor $L_{SD}/t_{Si}$ needed when $t_{Si}\ll L_{Sd}$. In order to compare with experiment, the spin accumulation is converted into a spin signal $V_{NL} (x) = P_{det}\,\Delta\mu(x)/2e$ that would be detected by a non-local spin detector contact placed at location $x$ and having a tunnel spin polarization $P_{det}$. From this, one obtains the spin signal per unit of injected current density $J$ (i.e., the spin-RA product $V_{NL}/J$).\\
\indent Our approach is different from the common practice\cite{note2}, in which the non-local spin transport data are analyzed without explicit integration over the width of the injector contact, considering the injector and detector to be line sources of infinitesimal width\cite{maekawa}. While this allows one to obtain a simple analytical expression for the magnitude of the non-local spin signal as a function of the distance between the injector and detector, the approximation is bound to fail when $W_{inj}$ is comparable to or larger than $L_{SD}$. Before we apply the model to the experimental data, we shall first describe the main predictions of the model with regard to the scaling of the spin accumulation as a function of $W_{inj}$, and examine to what extend the scaling behavior is captured by the approximation of the injector as a line.\\
\indent Fig. 3 displays the magnitude and the spatial profile of the spin accumulation as a function of $W_{inj}$, under the condition that the injector tunnel current {\em density} $J$ is kept constant. The following well-known features are to be noted. The $\Delta\mu$ has a maximum at the center of the injector and decays outside the injector region on both sides due to the isotropic spin diffusion in the Si channel, the decay being an exponential function $exp(-x/L_{SD})$ of the distance to the edge of the injector. For a very wide injector ($W_{inj} \gg L_{SD}$), the spin accumulation reaches a value of $2\,e\,J\,P_{inj}\,r_{ch}$, as it should\cite{jansensstreview}, but note that when the edge of the contact is approached, the spin accumulation is reduced (by exactly a factor of two, which can easily be understood; at the center of a very wide injector half of the spin accumulation is due to spins that were injected to the left and diffused to the center, the other half of the spins come from the right. At the edge, exactly one half is missing because no spins are injected outside the contact region). The decay near the edge occurs on the length scale of $L_{SD}$.\\
\indent The most important feature of the calculations is, however, that the maximum spin accumulation under the injector contact depends sensitively on the width of the injector, and more precisely, the spin accumulation is reduced when $W_{inj}$ becomes comparable to or smaller than the spin-diffusion length. The spin accumulation at the edge of the injector is shown in Fig. 3b (thick solid line). Two regimes can be identified. For $W_{inj} \gg L_{SD}$, the spin accumulation obtained with constant $J$ is independent of the contact width. For $W_{inj} \ll L_{SD}$, the spin accumulation decays linearly as a function of $W_{inj}$. Note that the scaling would be similar if we would plot $\Delta\mu/J$, irrespective of whether the current density $J$ or the total current $I$ is kept constant. If one were to plot $\Delta\mu/I$ instead, one would still have two regimes, however, $\Delta\mu/I$ would be constant for $W_{inj} \ll L_{SD}$ and decay linearly as a function of $W_{inj}$ for $W_{inj} \gg L_{SD}$. Notwithstanding, the calculations suggest a route to increase the spin accumulation and the resulting electrical spin signals, namely, by choosing an injector with a width that is comparable to or larger than the spin-diffusion length of the channel material.\\
\indent It is instructive to examine the range of validity of the commonly-used model in which the injector is approximated as a line\cite{maekawa}. Fig. 3a also displays the spin accumulation profiles for this case (dashed lines). The line injector model provides an accurate description of the spin accumulation profile when $W_{inj} \ll L_{SD}$. However, when $W_{inj}$ approaches $L_{SD}$, the profile under the contact is no longer properly described, and the maximum value of the spin accumulation at the contact center is significantly overestimated. However, the line injector model also overestimates the decay of $\Delta\mu$ between the contact center and the contact edges, so that the spin accumulation at the edge of the contact is reasonably well described up to larger contact width, i.e., up to $L_{SD}$ (see Fig. 3b). Note that we assumed that the line injector is placed at the center of the injector, which implies using the center-to-center distance in the exponential decay factor to describe the spin signals in non-local devices. We conclude that the line injector model describes non-local spin signals rather accurately for $W_{inj} \lesssim L_{SD}$, but for $W_{inj} \gtrsim L_{SD}$ one should use the exact numerical evaluation (expression (1)) that explicitly takes the injector width into account. Also, if one is interested in the spin accumulation under the injector contact, such as in a three-terminal measurement, one should not use the line injector model unless $W_{inj} \ll L_{SD}$.\\
\indent The scaling of the spin accumulation as a function of the size of the FM injector contact is an inherent part of the standard description of spin injection and diffusion in non-magnetic materials. Surprisingly, however, the scaling has never been experimentally tested. We will use the non-local spin transport devices with giant spin accumulation to first verify the predicted scaling of the spin signal and then extract the values of all the relevant spin-transport parameters from the data.

\begin{figure}[htb]
\centering
\includegraphics*[width=65mm]{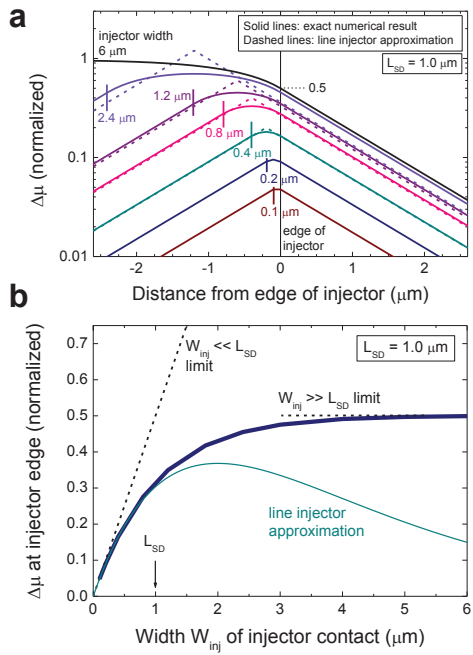}
\caption{Calculated spin accumulation. (a) Spatial profiles of the spin accumulation for different widths of the FM injector contact. Solid lines are the exact profile obtained from expression (\ref{eqn1}), whereas the dashed lines are for the approximation of the injector as a line of infinitesimal width, placed at the center of the injector (located between $x=0$ (right edge) and $x=-W_{inj}$ (left edge, indicated by the short colored vertical lines)). (b) Magnitude of the spin accumulation at the edge of the FM injector contact as a function of the contact width. The thick blue line is for the full numerical calculation using expression (\ref{eqn1}), whereas the thin green line is for the line injector approximation. For (a) as well as (b) we used L$_{SD}$ = 1.0 $\mu$m and the spin accumulation was normalized to the maximum value at the {\em center} of a very wide contact ($W_{inj} \gg L_{SD}$).}
\label{fig3new}
\end{figure}

\subsection{Role of injector width and spin-diffusion length} In order to test the calculations of the spin accumulation profile, non-local spin transport measurements were performed on devices with different contact dimensions and separations. Notably, for each device, data was collected for two configurations, using either the narrow FM strip as injector and the wider FM strip as detector, or vise versa, with an identical current density across the injector tunnel interface. Comparison reveals that the non-local spin signal is largest when the wider FM contact is used as injector, and smaller by a factor of 2 - 4 when the narrow FM is used as injector. This behavior is consistently observed for all devices investigated, examples of which are given in Fig. 4a and 4b. This feature, which has not been discussed before, is consistent with our calculations. In fact, eqn.(\ref{eqn1}) provides a good description of the magnitude of the non-local spin signal for all the devices with different $W_{inj}$ using $L_{SD}$ = 2.2 $\mu$m (see Fig. 4c). Note that besides material parameters (resistivity and thickness of the Si channel) the only other fitting parameters are the tunnel spin polarizations $P_{inj}$ and $P_{det}$. These determine the overall magnitude of the spin signals. We extract $P_{inj}\,P_{det}$ = 0.28, from which a tunnel spin polarization of 53 \% is obtained for the Fe/MgO/Si contacts, assuming that $P_{inj}$ = $P_{det}$ (which is valid if the detector and injector contact are identical and the current density is small enough so that the spin signal is linear in $J$, which was confirmed).\\
\indent It is stressed that the value of $L_{SD}$ not only controls the exponential decay of the spin accumulation as a function of distance from the edge of the injector, but also the scaling of the spin signal as a function of $W_{inj}$. Compared to the common procedure to determine the spin-diffusion length, in which only the gap between injector and detector is changed, our fitting procedure is more restrictive as it also includes the scaling with $W_{inj}$. In order to isolate the scaling, we compare the calculated spin signal at the edge of the injector contact to the values derived from the experimental data (Fig. 4d). The latter are obtained by compensating for the exponential decay between the edge of the injector and the center of the detector (i.e. by dividing out the factor $\exp(-x_{det}/L_{SD})$ with $x_{det}$ the location of the detector center). The result illustrates that indeed the spin accumulation is considerably reduced at small $W_{inj}$. For the smallest size (0.4 $\mu$m) used here, the spin accumulation is a factor of 5 smaller than the maximum value that can be obtained for a very wide injector with $W_{inj} \gg L_{SD}$.

\begin{figure}[htb]
\centering
\includegraphics*[width=76mm]{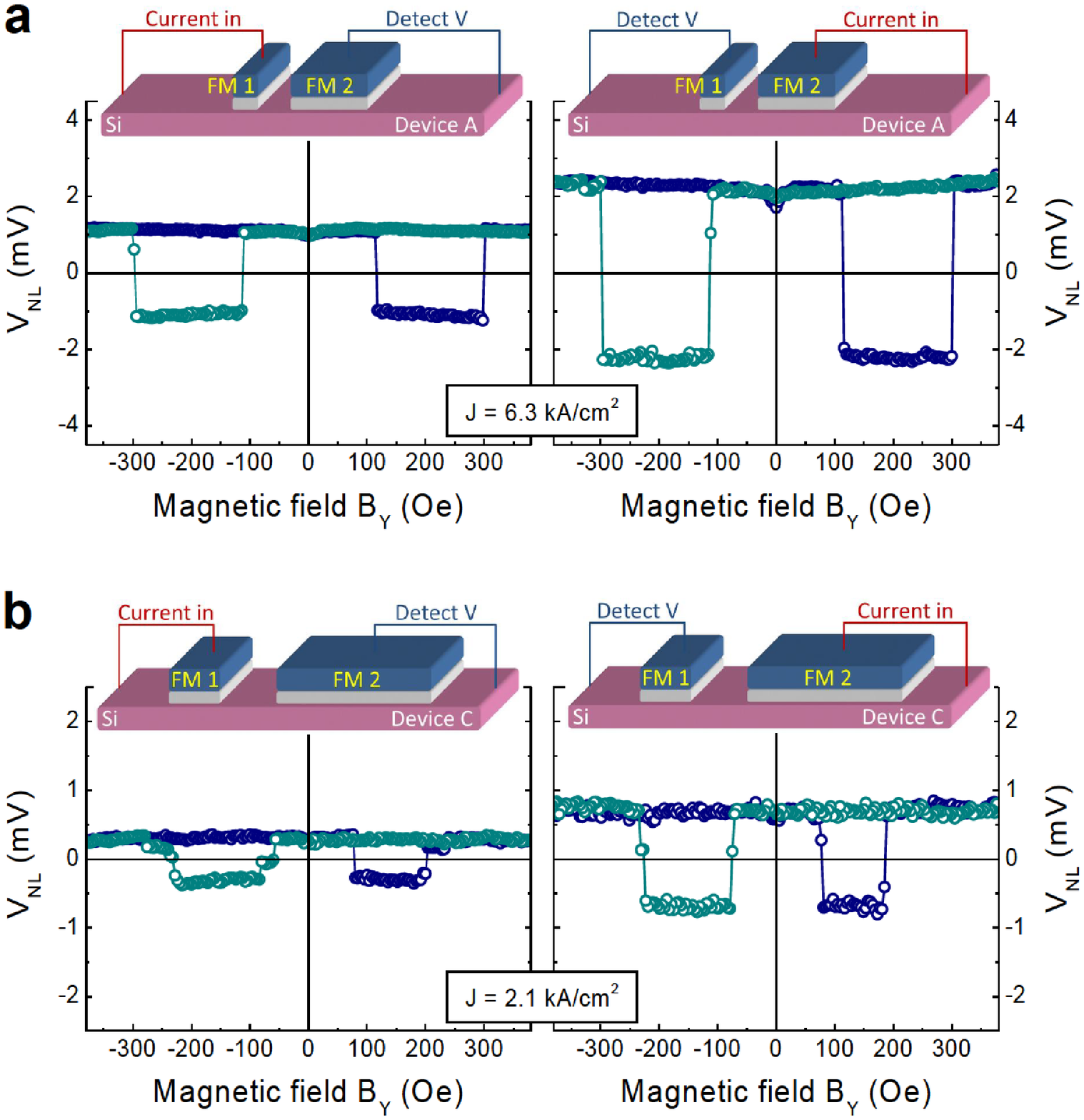}\hspace*{5mm}\includegraphics*[width=55mm]{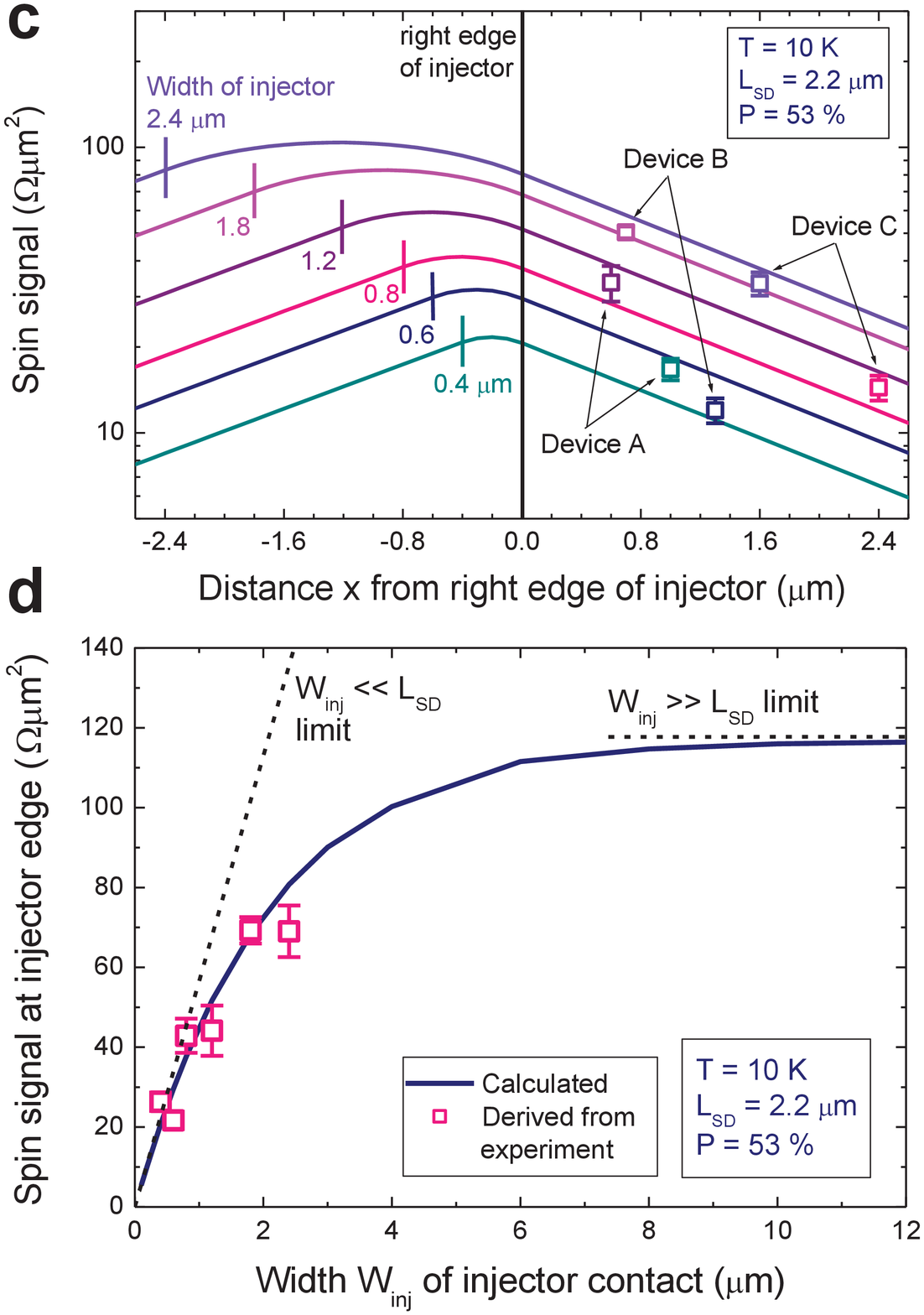}
\caption{Calculated spin signal and experimental data for different injector contact width. (a) Non-local spin-valve signals at 10 K for device A at J = +6.3 kA/cm$^2$ using either the narrow FM strip as injector and the wider FM strip as non-local detector (left panel, $I$=+1mA), or vise versa (right panel, $I$=+3mA), as indicated by the schematic diagrams of the measurement configuration. (b) the same for device C at J = +2.1 kA/cm$^2$ (left panel $I$=+0.66mA, right panel $I$=+2mA). (c) Calculated spin signal versus position produced by the spin accumulation in the Si for different widths of the FM injector contact (solid lines), and experimental data (symbols) for devices A, B and C obtained in two configurations (either using the narrow FM strip as injector and the wider FM strip as detector, or vise versa). The injector is located between $x=0$ (right edge) and $x=-W_{inj}$ (left edge, indicated by the short colored vertical lines). The experimental data is compared to the calculated spin signal at the center of the detector, which is located at $x>0$. The best agreement with the data is obtained using L$_{SD}$ = 2.2 $\mu$m and P = 53 \% in the calculation. (d) Calculated spin signal (solid line) at the right edge of the injector ($x=0$) versus width of the injector contact, compared to the values (symbols) derived from the experimental data by compensating for the exponential decay between $x=0$ and the center of the detector.}
\label{fig4}
\end{figure}

\indent The spin lifetime of the Si is extracted from non-local Hanle measurements on different devices using spin injector contacts of different widths (Fig. 5, T = 10 K). Besides the known geometrical factors, the shape of the Hanle curve depends on the combination of $\tau_s$ and $L_{SD}$, while the amplitude of the Hanle signal is set by the value of $L_{SD}$. It is therefore customary to use both as fitting parameters. We used a different procedure that makes the extracted value of $\tau_s$ insensitive to any variations in the amplitude of the experimental Hanle signal. These are always present and most likely arise from variations of the tunnel spin polarization of the contacts or deviations from a perfect P or AP magnetization alignment during the Hanle measurement. Therefore, we set the value of $L_{SD}$ to 2.2 $\mu$m as determined above, apply a scaling factor to adjust the signal amplitude if needed, which then leaves $\tau_s$ as the only parameter to fit the shape of the Hanle curve. The Hanle curves are then well described by the numerical evaluation of eqn. (\ref{eqn1}) and a good fit is obtained by using the same value of $\tau_s$ of 18 ns for all the configurations with different injector and detector widths and spacing.\\
\indent We conclude that the spin-transport model, that explicitly takes the width of the injector contact into account, provides an adequate and consistent description of all the non-local spin-transport data with reasonable values of the extracted parameters. Based on the analysis, we attribute the giant spin accumulation to two factors: (i) the large tunnel spin polarization of the Fe/MgO contacts to the Si (53 \% at 10 K), and (ii) the spin density enhancement achieved by using a spin injector with a size comparable to or larger than the spin-diffusion length of the Si.

\begin{figure}[htb]
\centering
\includegraphics*[width=105mm]{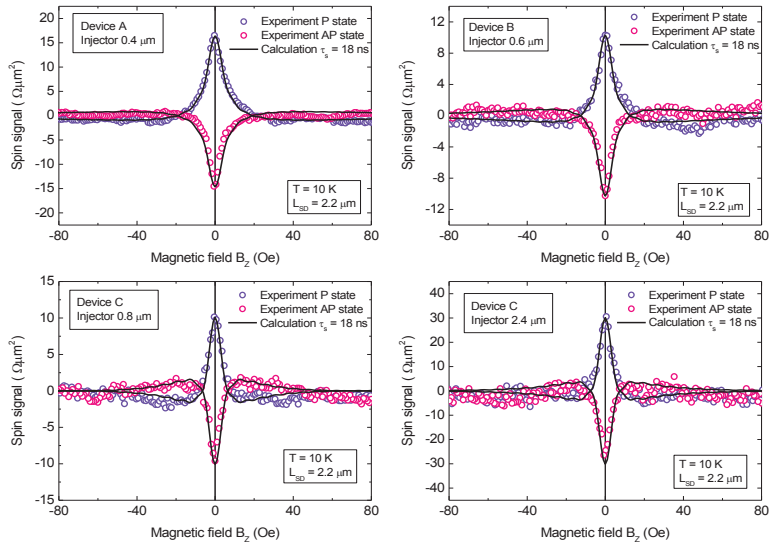}
\caption{Non-local Hanle measurements and extracted spin lifetime. Non-local Hanle measurements for parallel (blue open circles) and antiparallel (pink open circles) orientation of the magnetization of the injector and the detector, for different devices and width of the injector, as indicated. The solid black lines correspond to the numerically calculated Hanle signal at the center of the detector using $\tau_s$ = 18 ns, which simultaneously provides a good fit for all data
sets. T = 10 K.}
\label{fig5}
\end{figure}

\subsection{Non-local spin transport at room temperature} Since a giant spin accumulation persists up to room temperature, we analyze the room temperature spin signals in more detail in order to extract the relevant parameters. Measurement are performed on devices with different dimensions of the contacts, and for each device data was collected for two configurations, using either the narrow or the wider FM strip as injector. The data for device A is presented in Fig. 6a. Clear non-local spin-valve and Hanle signals with consistent magnitudes are observed for both configurations, but the spin signals are about 4 times larger when the wider FM strip is used as injector. Also at room temperature, the magnitude of the non-local spin signals for different $W_{inj}$ is well described by the numerical calculations of the spin accumulation profile (Fig. 6b), for L$_{SD}$ = 1.0 $\mu$m and $P$ = 18\%. Because $L_{SD}$ is smaller at 300 K, the maximum spin accumulation does not depend very much on $W_{inj}$ for contact widths of 0.8, 1.2 and 2.4 $\mu$m that are comparable to or larger than $L_{SD}$, but a significant reduction is still present for $W_{inj}$ = 0.4 $\mu$m. From the fit of the Hanle data a spin lifetime of $\tau_s$ = 2.5 ns at 300 K is obtained.\\
\indent Finally, we compare our data to previous work on non-local spin-transport devices, specifically, by Suzuki et al. \cite{suzukiAPEX2011} who first reported non-local spin transport in Si devices up to room temperature, and by Ishikawa et al. \cite{ishikawaPRB2017}, who only very recently published data with, to the best of our knowledge, the largest non-local spin signals for degenerately-doped Si to date. We compare the published experimental data \cite{suzukiAPEX2011,ishikawaPRB2017} of the spin signals converted into a spin-RA product (units of $\Omega\mu$m$^2$), instead of previously extracted values of $\Delta\mu$ or $P$, in order to make the comparison insensitive to differences in the theoretical analysis used to extract parameters from the data. The comparison, displayed in Fig. 6c, reveals that the spin signals reported here are significantly larger, by 2 - 3 orders of magnitude, showing the significant improvement achieved. Most of this improvement (1-2 orders of magnitude) is due to the larger tunnel spin polarization of the Fe/MgO contacts, as can be deduced from a comparison of our data for the device with the 0.4 $\mu$m wide injector with the data of Ref. \cite{suzukiAPEX2011,ishikawaPRB2017}, in which an injector with comparable size was used. The increase of the contact width provides another, yet more modest, increase of the spin signal (by a factor of 3-4). It is difficult to isolate the exact origin(s) of the improved device performance, as there can be many reasons why a large tunnel spin polarization, as expected for the Fe/MgO system, was obtained here but not in previous reports\cite{suzukiAPEX2011,shiraishi,sasaki,toshibanonlocal,ishikawaPRB2017,toshiba2017}. The differences in the fabrication processes include (i) the way the Si surface was prepared (chemical cleaning process, annealing temperature, type of surface reconstruction obtained), (ii) the growth of the MgO (deposition temperature, MgO thickness), (iii) the growth of the ferromagnetic electrode (deposition method, growth temperature), and (iv) the etching process used to define the tunnel contact area (including possible damage at the contact edges and edge leakage currents).\\
\indent It is notable that the spin signal at room temperature is within about a factor of 2 from the largest non-local spin signals ever observed, in devices with a graphene channel\cite{kawakamiprl,kawakamireview}. We observe that the maximum spin signal decays by roughly a factor of 12 between 10 and 300 K. For the most part, this decay originates from the decay of the tunnel spin polarization (factor of (53/18)$^2$ = 8.6). The rest is due to the variation of $L_{SD}$ with some compensation due to the factor of 2 increase of the resistivity of the Si between 10 and 300 K (see Appendix A). This suggests that further improvement of the spin signal is most likely to come from optimizing the tunnel contacts to achieve even higher tunnel spin polarization and/or reduce the decay of the polarization with temperature.

\begin{figure}[htb]
\centering
\includegraphics*[width=111mm]{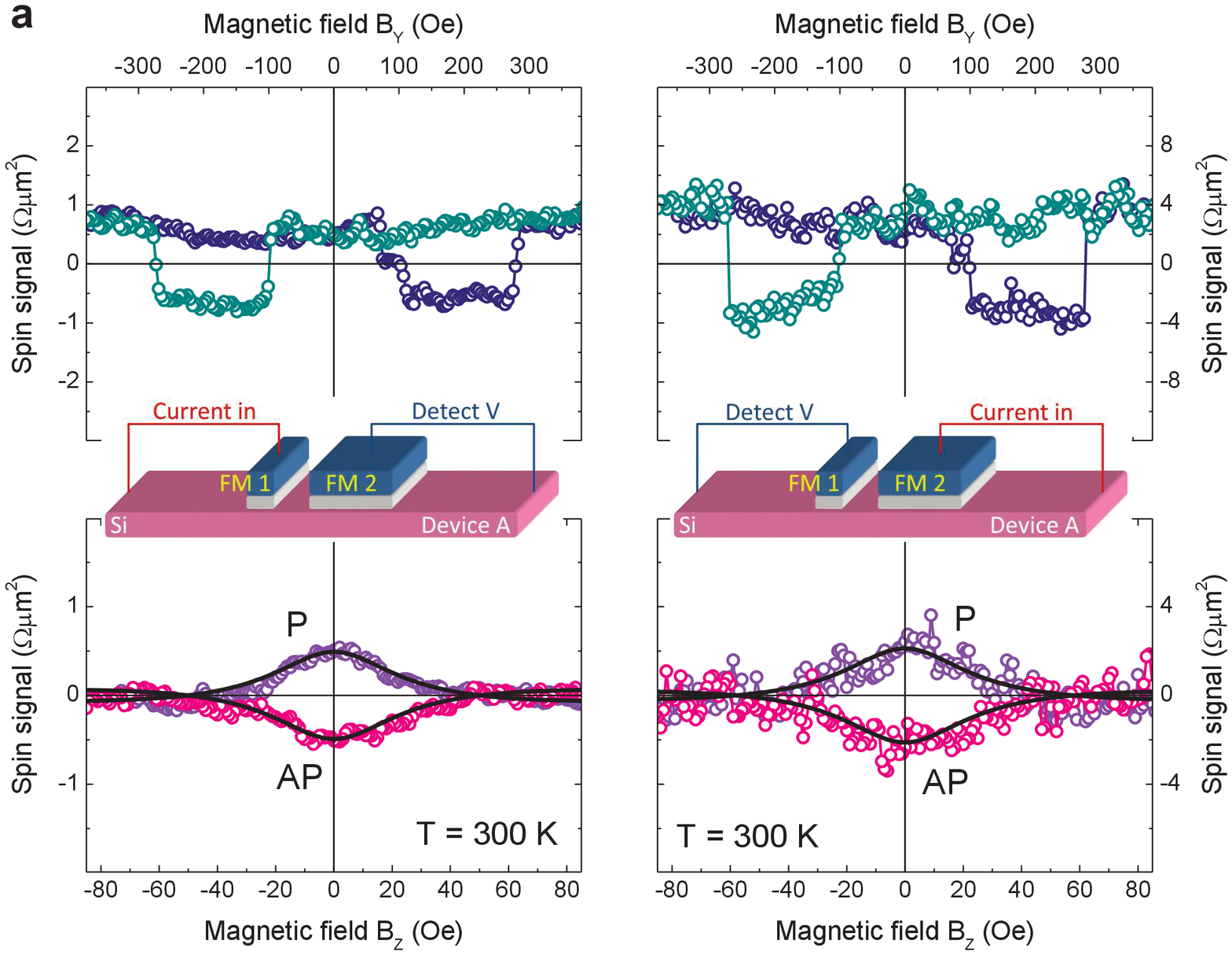}\hspace*{5mm}\includegraphics*[width=65mm]{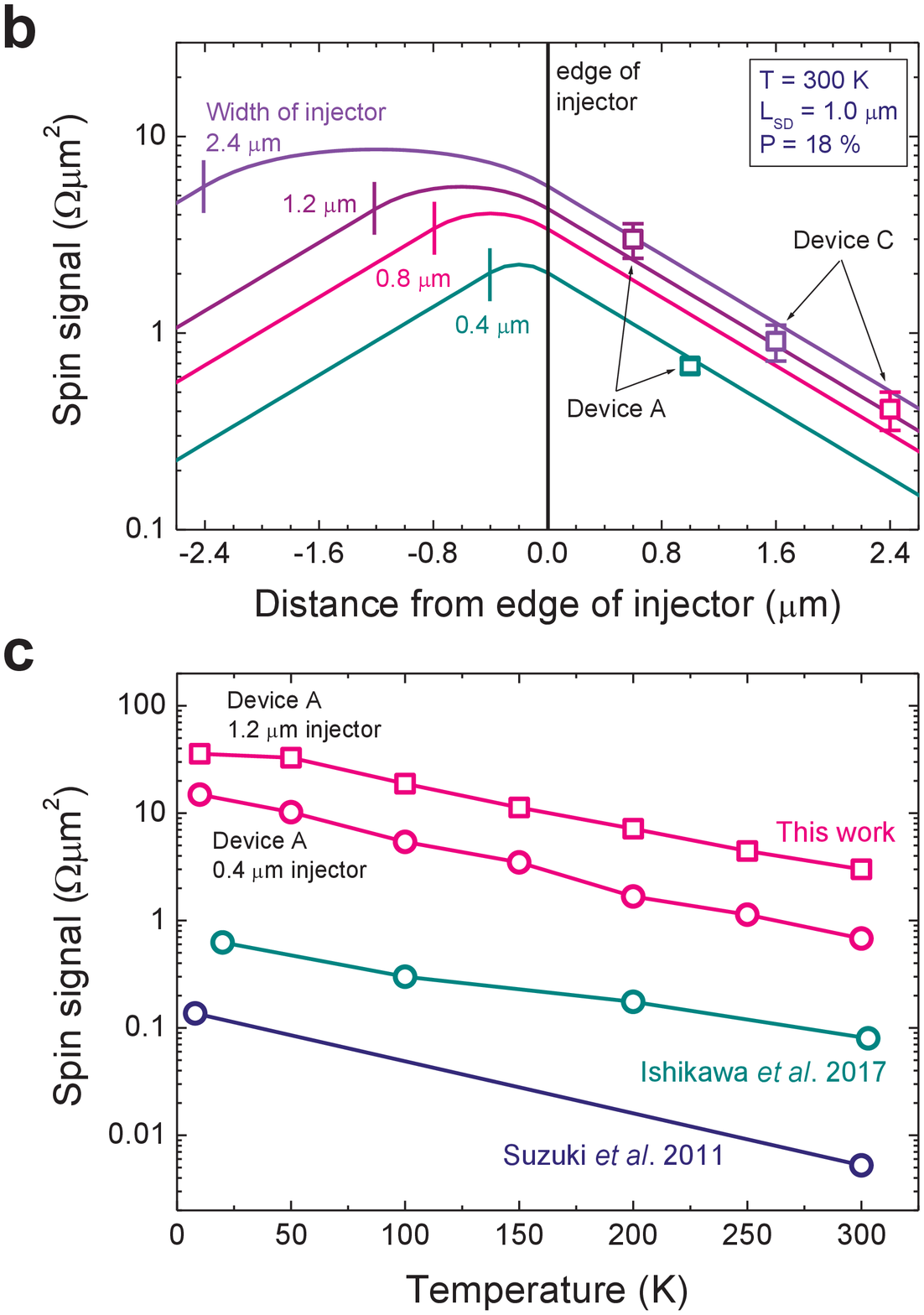}
\caption{Spin transport in Si at room temperature. (a) Non-local spin signals at T = 300 K for device A using either the narrow FM strip as injector and the wider FM strip as non-local detector (left panels, J = +7.5 kA/cm$^2$, $I$ = +1.2 mA), or vise versa (right panels, J = +4.2 kA/cm$^2$, $I$ = +2 mA), as indicated by the schematic diagrams of the measurement configuration. The top panels display the spin signal for the spin-valve geometry, the bottom panels for the Hanle geometry. The fits to the Hanle data (solid black lines) are obtained using $\tau_s$ = 2.5 ns. Note the factor of 4 difference in the vertical scale between the left and right panels. (b) Spin signal at 300 K calculated as a function of position in the Si channel for different widths of the FM injector contact (solid lines), and experimental data (symbols) for devices A and C obtained in two configurations (either using the narrow FM strip as injector and the wider FM strip as detector, or vise versa). The best agreement with the data is obtained for L$_{SD}$ = 1.0 $\mu$m and P = 18 \%. (c) Non-local spin signals versus temperature for device A for two configurations (pink symbols) together with previous data taken from Ref.\protect\onlinecite{suzukiAPEX2011} (dark blue symbols) and Ref.\protect\onlinecite{ishikawaPRB2017} (dark green symbols).}
\label{fig6}
\end{figure}

\section{Summary}
\noindent Using non-local spin transport devices, it was demonstrated that it is possible to create a giant spin accumulation in degenerately doped silicon, with the spin splitting reaching values as large as 13 meV at 10 K and 3.5 meV at room temperature. Numerical evaluation of a spin-transport model that explicitly takes the width of the injector contact into account provides an adequate and consistent description of all the non-local spin-transport data with reasonable values of the extracted parameters. Based on the analysis, we attribute the giant spin accumulation to two factors: (i) the large tunnel spin polarization of the Fe/MgO contacts to the Si (53 \% at 10 K and 18 \% at 300 K), and (ii) the spin density enhancement achieved by using a spin injector with a size comparable to or larger than the spin-diffusion length of the Si.\\

\begin{appendix}
\section{Experimental details} The Fe/MgO tunnel contacts were grown by molecular beam epitaxy (MBE) on a 70 nm-thick phosphorous-doped n-type Si(001) channel on an undoped Si substrate. The carrier density of the Si channel determined via the Hall effect was 2.7 $\times$ 10$^{19}$ cm$^{-3}$ at 10 K and 1.6 $\times$ 10$^{19}$ cm$^{-3}$ at 300 K. The measurements of the Si resistivity in the Van der Pauw geometry yielded 1.3 m$\Omega$cm at 10 K and 2.6 m$\Omega$cm at 300 K. Prior to the deposition of the FM tunnel contact, the Si substrate was cleaned using a so-called RCA process that includes treatments with alkaline (NH$_4$OH:H$_2$O$_2$:H$_2$O) and acidic (HCl:H$_2$O$_2$:H$_2$O) hydrogen peroxide solutions. This ensures the removal of organic and metallic contaminants and creates a smooth surface. The substrate was then etched in dilute hydrofluoric acid (2\%) and rinsed with deionized water to remove the oxide and produce a hydrogen-terminated surface. After introduction into the MBE system having a base pressure in the high 10$^{-10}$ Torr range, the substrate was annealed at 700 $^{\circ}$C for 10 min to desorb the hydrogen and obtain a clean Si surface. Subsequently, a 2 nm-thick MgO layer and a 10 nm-thick Fe layer were deposited at 300 $^{\circ}$C and 200 $^{\circ}$C, respectively. To avoid the oxidation of the Fe layer, the sample was covered by a 20-nm-thick Au capping layer. The four-terminal Si lateral devices were prepared by standard micro-fabrication techniques (e-beam lithography, Ar-milling, SiO$_2$ sputtering) and consist of two FM electrodes and two nonmagnetic Au/Ti reference electrodes contacting the Si channel, that is patterned into a 50 $\mu$m wide strip. A single chip contains many electrically isolated devices with various dimensions of the electrode strips and their spacing.\\
The charge transport properties of the Fe/MgO/Si tunnel contacts were investigated by measuring the current density versus voltage (J-V) characteristics in a three-terminal configuration. Positive current corresponds to electrons flowing from the ferromagnet into the Si. The resistance area product (RA) of the junctions was found to be in the range of a few times 10 k$\Omega\mu$m$^2$ or higher. This is significantly larger than the effective spin resistance\cite{jansensstreview} of the Si channel $r_{ch}\,(1-P^2)\,\sim\,$640 $\Omega\mu$m$^2$ at 10 K, which ensures that the spin accumulation is not reduced by back-flow of the spins from the Si into the FM contacts (also referred to as spin absorption or conductivity mismatch). For the current densities used in the non-local spin-transport measurements (+2.1 to +12.5 kA/cm$^2$), the voltage across the injector contact is in the range of +0.8 to +1.7 V. The current density $J$ is defined using the lateral area of the injector FM tunnel contact.
\end{appendix}

\begin{acknowledgments}
This work was supported by the Grant-in-Aid for Scientific Research on Innovative Areas, “Nano Spin Conversion Science” (Grants No.26103002 and 26103003).\\
\end{acknowledgments}



\begin{thebibliography}{99}
\bibitem{note1} The factor of 2 in eqn. (\ref{eqn1}) is needed because $\Delta\mu$ is defined as $\mu^{\uparrow}-\mu^{\downarrow}$, instead of half
of this, as is sometimes also done. Also note that the factor $1/\tau_s$ in the integration ensures that the maximum spin signal in the center of a
very large contact ($W_y\gg L_{sd}$) is equal to spin-RA = $P_{inj}\,P_{det}\,r_{ch}$, as it should be.
\bibitem{note2} In a few cases, involving channel materials other than Si, authors have evaluated the non-local spin signals by either a Monte-Carlo approach
or by a numerically solving the spin-transport equations using finite-element methods, in which case the width of the injector contact is automatically taken into account.
\end{thebibliography}
\end{document}